\documentclass[11pt]{article}
\usepackage{moriond}
\usepackage{graphicx}
\usepackage{floatflt}
\usepackage{xspace}
\usepackage{amsmath}
\usepackage{amssymb}
\usepackage{url}

\newcommand{\order}[1]{{\cal O}\left(#1\right)}
\newcommand{\MeV}{\,\mathrm{MeV}} \newcommand{\GeV}{\,\mathrm{GeV}}
\newcommand{\TeV}{\,\mathrm{TeV}}
 \newcommand{\tw}{\textwidth}
 \newcommand{\Dzero}{D\O\xspace}
\newcommand{\cs}{$\mspace{3.0mu}$}

\def\be{\begin{equation}}
\def\ee{\end{equation}}
\def\bea{\begin{eqnarray}}
\def\eea{\end{eqnarray}}

\begin{document}
\vspace*{4cm}
\title{MORIOND 2009, QCD AND HIGH
  ENERGY INTERACTIONS:  THEORY SUMMARY }

\author{GAVIN P. SALAM}

\address{LPTHE, UPMC Univ. Paris 6, CNRS UMR 7589, Paris, France}

\maketitle\abstracts{
  These proceedings provide a brief summary of the theoretical topics that were
  covered at Moriond QCD 2009, including non-perturbative QCD,
  perturbative QCD at colliders, a small component of physics beyond
  the standard model and heavy-ion collisions.
}

\section{Introduction}

Of the $\order{100}$ talks that were given at this year's ``Moriond
QCD'', about one third were theoretical.\footnote{%
  Two thirds of those were in the afternoon, which means that at
  roughly 95\% confidence level, we can rule out the hypothesis that
  the organisers were equally likely to assign theory talks to morning
  and afternoon sessions.
  In contrast with the Tevatron's 95\% exclusion limits on the
  standard-model Higgs boson with
  $160<m_H<170\GeV/c$,\cite{HiggsExclusion} it is, however, unclear
  quite what we learn from this!
}
As usual with the Moriond conference, the range of topics covered was
rather broad, and the logic that I will follow in discussing them will
be to progress in the total energy that is involved --- that will take
us from non-perturbative QCD, through perturbative QCD and the
data-theory interface at high-energy colliders, to topics beyond the
standard model, and finally to heavy-ion collisions.

\section{Non (or barely) perturbative QCD}
\label{sec:lattice}

There are many reasons for investigating non-perturbative QCD. One
good one is that it's responsible for most of the nucleon mass and
correspondingly for most of the visible mass in the universe.
A more pragmatic reason is that flavour physics is usually done with
hadrons, and our understanding of their non-perturbative dynamics is
one of the limiting factors in the extraction of CKM matrix entries
and new-physics constraints.

A powerful tool for handling non-perturbative QCD is to
simulate it on the lattice. 
A recurrent issue for lattice QCD is the reliable handling of
systematic errors, for example the dependence on the lattice spacing,
the matching of lattice gauge theory to continuum QCD, finite-volume
effects, and the treatment of light quarks, and the discussion of
these issues was a common theme to the lattice talks at Moriond '09.

Three light-quark treatments were discussed.
Staggered fermions are the easiest to treat from a computational point
of view, but this comes at a price: while the predictions agree with
experimental results, it is not clear whether the staggered-fermion
formulation is theoretically equivalent to QCD.
Wilson fermions and domain-wall fermions are both OK from this point
of view, but they are also more expensive computationally, especially
the domain-wall fermions,
which are those with the cleanest chiral ($m_q \to 0$) limit.

\begin{figure}
  \centering
  \includegraphics[height=0.23\tw,width=0.32\tw]{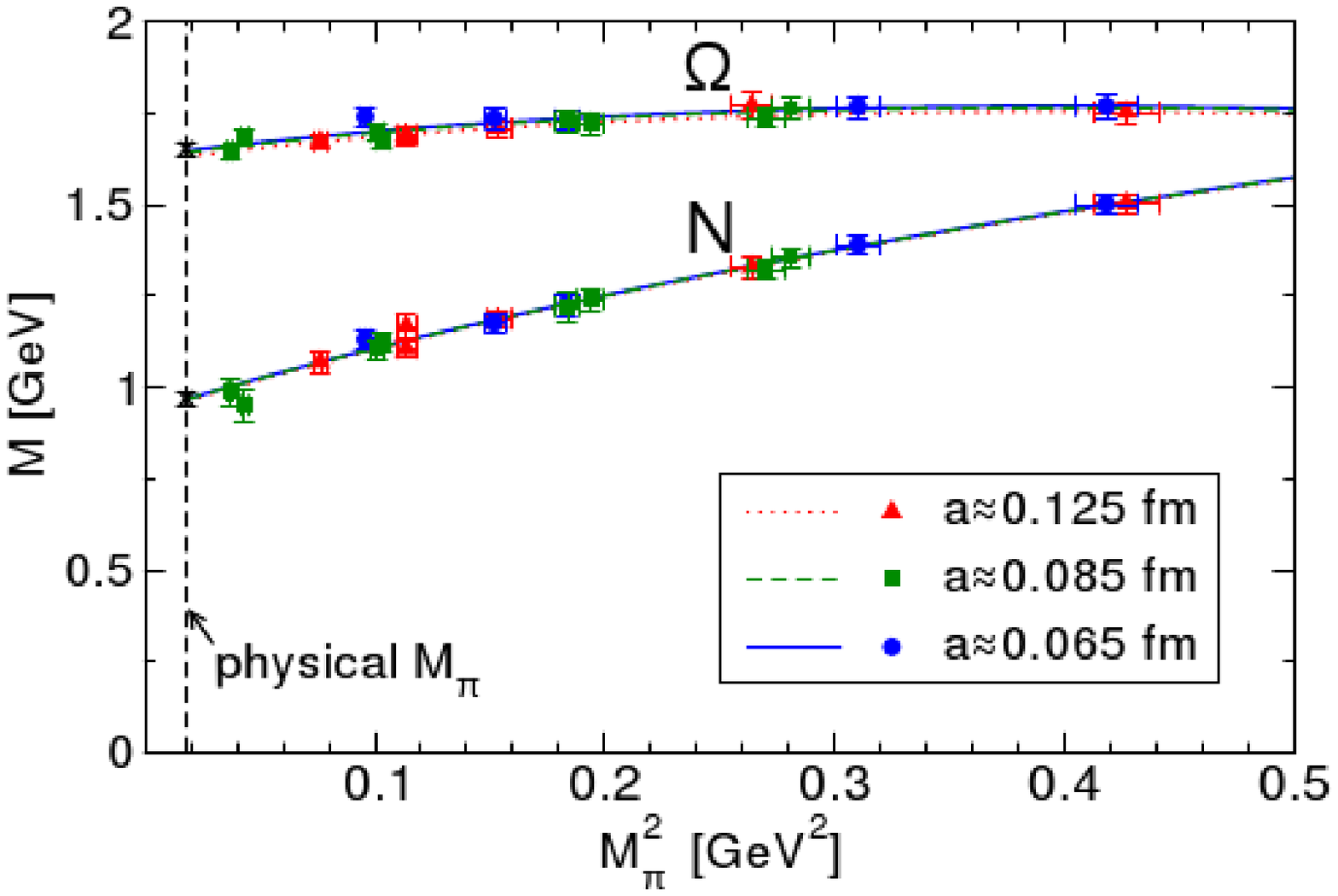}
  \includegraphics[height=0.23\tw,width=0.32\tw]{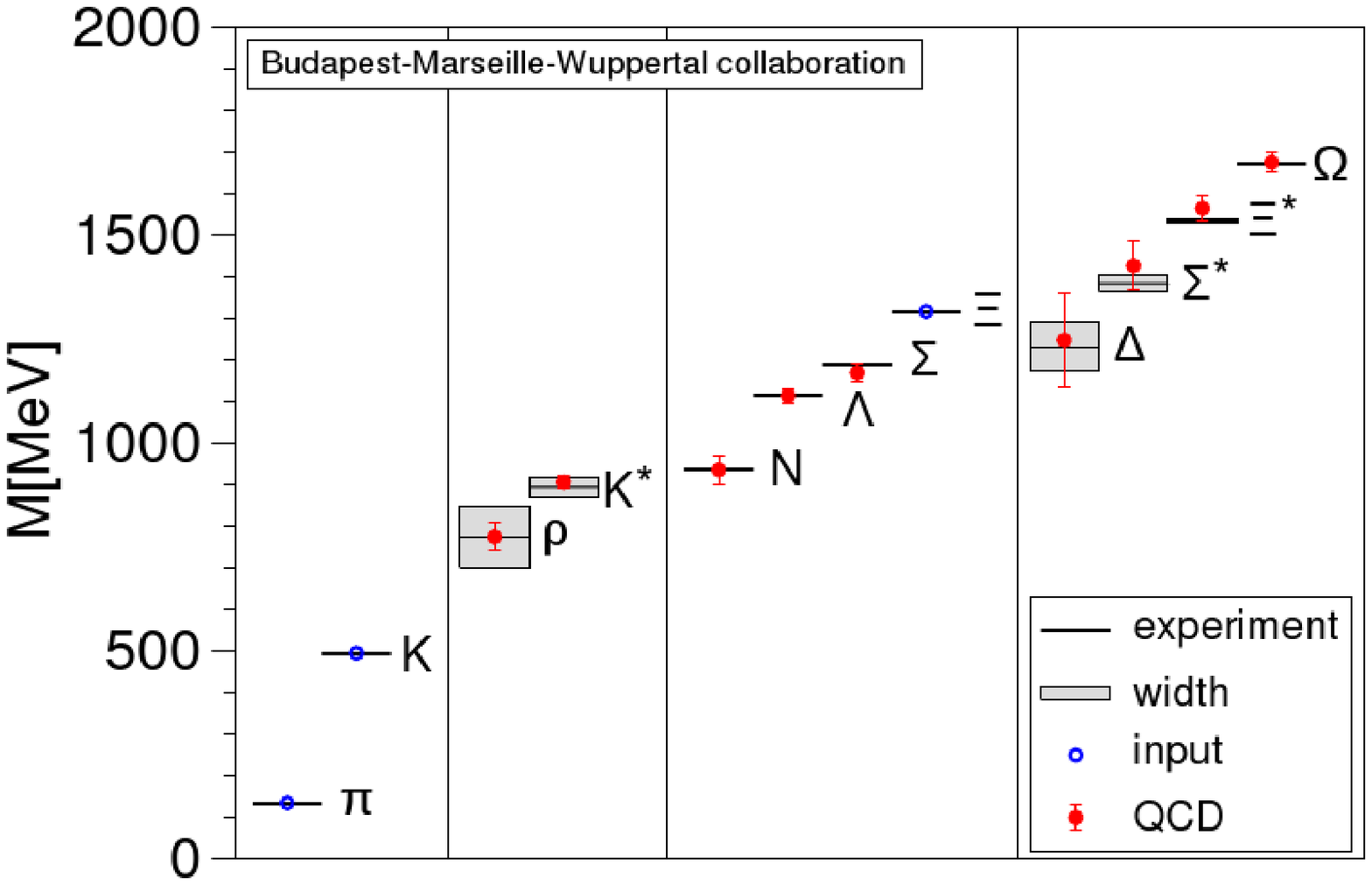}\;\;
  \includegraphics[height=0.23\tw,width=0.32\tw]{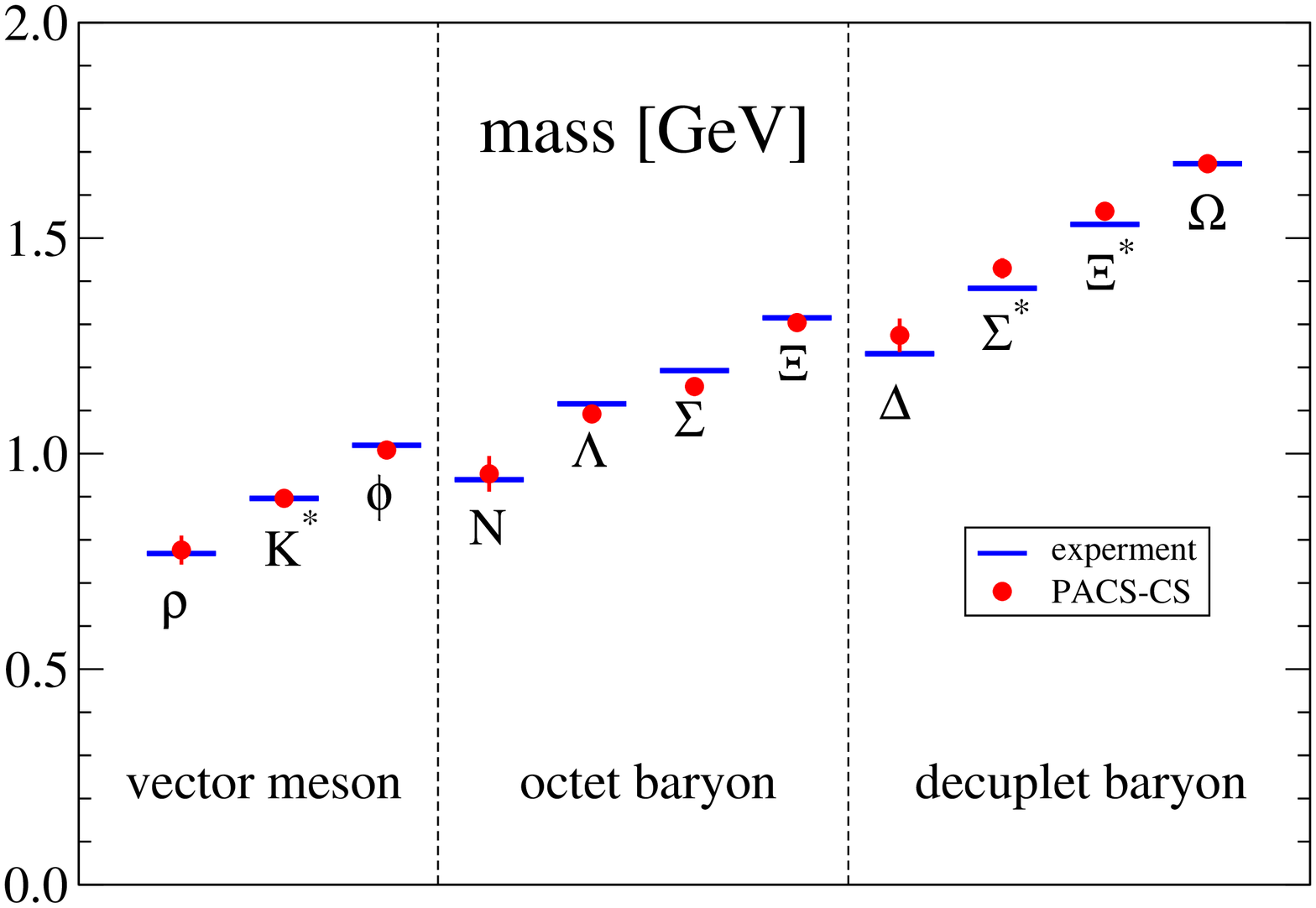}
  \caption[dummy]{Left: dependence of the $\Omega$ and nucleon masses
    on the pion mass and lattice spacing from the BMW
    collaboration;\cite{FodorProc} middle: final results for the
    hadron mass spectrum from the BMW collaboration;\cite{FodorProc}
    right: results for the spectrum from the PACS-CS
    collaboration based on a linear extrapolation from the region $156\MeV
    < m_\pi <410\MeV$.\cite{KuramashiProc}}
  \label{fig:lattice-nucleon-mass}
\end{figure}

Figure~\ref{fig:lattice-nucleon-mass} (left) illustrates results from two
talks about the hadron mass spectrum. The left-hand plot, presented by
Fodor\cs\cite{FodorProc} for the ``BMW'' collaboration,\cite{Durr:2008zz} shows how the lattice
calculation of the $\Omega$ and nucleon (N) masses depends on the
squared pion mass (horizontal axis), i.e.\ the approach to the correct
$u$ and $d$-quark masses, and on the lattice spacing $a$ (differently
coloured points), together with a fit that provides an extrapolation
to the physical light-quark masses.\cite{Durr:2008zz} 
The corresponding results for the hadron spectrum are shown in the
middle plot (lines and bands are experimental masses and widths, the
points are the lattice result), with remarkable agreement for all the
hadrons. 
This was presented as the first lattice-calculation of the baryon mass
to have full control of uncertainties.
Related results were presented by Kuramashi\cs\cite{KuramashiProc} for
the PACS-CS collaboration (right-hand figure).\cite{Ishikawa:2009vc}
He, however, argued that for a fully controlled calculation one should
carry out simulation directly with the physical light-quark masses
(currently in progress).
This is to avoid the extrapolation that is required in the BMW results
and
whose validity was the subject of debate during the conference.
%
%
%
%
Though there does not yet seem to a be a universal consensus within
the lattice community as to whether the hadron spectrum is now
reliably calculated, 
it is to be expected that clarification on the remaining issues will
be forthcoming in the near future.
Given the 35 years' work on the subject, that is a
major accomplishment, both in fundamental terms, and because it helps
provide confidence when using lattice results for observables for
which we don't already know the answer.

\begin{floatingfigure}[r]{0.45\textwidth}
  \centering
  \includegraphics[width=0.4\textwidth]{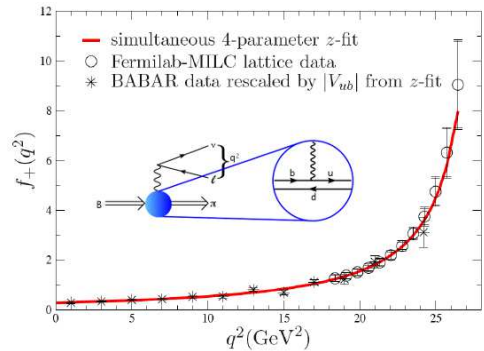}
  \caption[]{$B\to \pi \ell\nu$ form-factor as a function of $q^2$, the
    invariant mass of the $\ell \nu$ system, comparing experimental
    measurements\cs\cite{Aubert:2006px} and lattice
    calculations.\cite{Bailey:2008wp}}
\label{fig:vandewater}
\end{floatingfigure}
The use of lattice calculation to obtain information that we don't
know was illustrated in two other talks. 
Izubuchi,\cite{IzubchiProc} representing the RBC/UKQCD collaboration,
showed numerous results, including hadronic matrix element
computations, and determinations of the up- and down-quark mass
difference, and emphasised the value of the continuum chiral behaviour
that is characteristic of the domain-wall fermions that were used.

Van de Water,\cite{VandeWaterProc} for the Fermilab Lattice and MILC
collaborations (staggered fermions), discussed results for $B$-mesons
and their relation with the determination of CKM matrix elements.
Fig.~\ref{fig:vandewater} shows how data\cite{Aubert:2006px} for the
$B\to\pi \ell\nu$
form factor from BABAR and from the lattice calculation\cite{Bailey:2008wp} have the same
shape in the region of overlapping $q^2$ values, helping to provide
confidence in the lattice calculation and the extraction of $V_{ub}$.
The resulting value for $V_{ub} =  (3.38\pm0.36)\times10^{-3}$, while it has
errors ($11\%$) that are slightly larger than inclusive determinations
(7--8\%),\cite{InclusiveVub} is in better agreement with
unitarity triangle analyses $V_{ub} = (3.46\pm0.16)\times10^{-3}$.\cite{CKMVub}

With other observables it may currently be harder for lattice QCD to
provide definitive predictions. One context where this was discussed,
by Penin,\cite{PeninProc} concerned the mass of the recently
discovered\cs\cite{BABAR:2008vj} $\eta_b$, specifically the hyperfine
mass splitting, measured to be $E_{hfs} \equiv M(\Upsilon(1S)) -
M(\eta_b)=71.4\pm2.7^{+2.3}_{-3.1}$\,MeV.
In the charmonium system, the experimental value is well reproduced by
a perturbative calculation,\cite{Kniehl:2003ap} but this is not the
case for the $\eta_b$, where the prediction was $E_{hfs} =
39\pm11\mathrm{(th)}^{+9}_{-8}(\delta \alpha_s)$\,MeV.
The lattice prediction\cs\cite{Gray:2005ur} is closer to the experimental
result, however Penin argued that the lattice's coarse spacing
relative to the inverse $b$-quark mass implies
substantial additional corrections  ($\sim -20$\,MeV), which would bring it
into accord with the perturbative result.
This leaves an interesting puzzle, perhaps to be resolved at a future
Moriond!

Another context where the question of the lattice's predictive ability
naturally arose was the talk by Swanson\cs\cite{SwansonProc} about exotic
hadronic states. An example that was particularly interesting (though
it is unclear if it truly exists) was the $Z^\pm(4430)$, which decays
to $\pi^\pm \psi'$ and so would call for either a tetraquark or a
molecular interpretation.
As progress in lattice calculations continues, one can only look
forward to the day when they will be able to shed light on the
existence and structures of the numerous X,Y,Z resonances that are
currently being seen by the experiments.

\section{Perturbative QCD predictions}

Perturbative QCD (pQCD) inevitably ``happens'' at HERA, the Tevatron
and LHC. Backgrounds to possible new physics all involve a QCD
component, and more often than not, possible signals either involve
QCD directly (e.g.\ because a new particle decays to quarks) or are
affected, e.g.\ by pQCD initial-state radiation.

\subsection{NLO}
\label{sec:NLO} 

\begin{figure}
  \centering
  \includegraphics[width=0.7\tw]{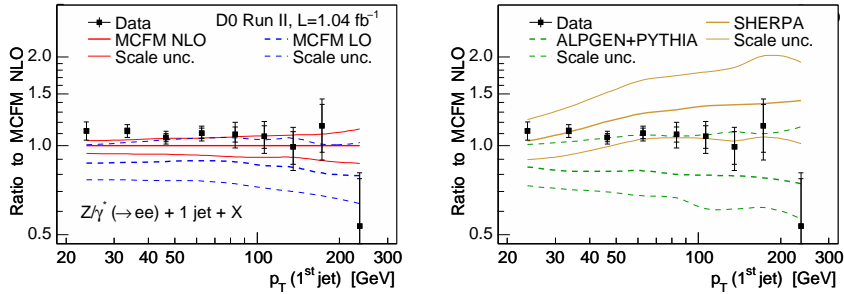}
  \caption[]{Ratio of measured Z+jet cross
    sections\cs\cite{NilsenProc,Abazov:2009av} to the NLO
    prediction\cs\cite{Campbell:2002tg} including its scale
    uncertainty, left, and compared to LO matrix-element plus
    parton-shower calculations,\cite{Hoche:2006ph} right (adapted from
    ref.~\cite{Abazov:2009av}).}
  \label{fig:nilsen-vjet}
\end{figure}
A number of the pQCD results presented here related to next-to-leading
order (NLO) calculations. The importance of NLO predictions was nicely
illustrated by Nilsen (for the \Dzero
collaboration\cs\cite{NilsenProc,Abazov:2009av}) in his comparisons of
data for the $Z$+jet cross-section to LO (matrix-element +
parton-shower\cs\cite{Hoche:2006ph}) and NLO
predictions,\cite{Campbell:2002tg} fig.~\ref{fig:nilsen-vjet}. It is
clear that it is only at NLO that one has a reliable prediction.
The usefulness of NLO predictions has led to the establishment of the
so-called Les Houches wish-list of important processes to calculate at
NLO, and this guides much of the current work on the
subject.\cite{Bern:2008ef}

The NLO results reported here can be split into two categories: those
that push traditional Feynman-diagram based methods to their limit
(J\"ager, Weinzierl), and those based on ``unitarity'' (Melnikov,
Ma\^\i tre) for the 1-loop part of the computation.

J\"ager\cs\cite{JaegerProc} discussed $pp \to VVjj$, via vector-boson
fusion (VBF), which is an important background to Higgs production via
VBF, and of interest also for studying WW scattering. She showed that
the NLO corrections\cs\cite{Jager:2006zc} are modest, and lead to small
scale-dependence in the final predictions, and illustrated how this
might facilitate the identification of new physics signal in the gauge
sector.
Weinzierl\cs\cite{WeinzierlProc} discussed $p\bar p \to t\bar t j$
production,\cite{Dittmaier:2007wz} one of the last uncalculated $2\to
3$ ``Les Houches'' processes, whose complexity stems from the
significant number (450) of loop diagrams and the fact that they
contain a mass scale, $m_t$.
One of the interests of $t\bar t j$ production is that its LO
contribution is the first order of $t\bar t$ production that shows an
asymmetry between the $t$ and $\bar t$ directions (jets are
preferentially emitted when the $t$ goes in the direction opposite to
the $p$). Curiously this asymmetry is largely washed out by higher
order corrections, an effect that calls for a physical explanation.

The bottleneck in NLO calculations for a $2\to n$ process is the
$2+n$-leg loop calculation, whose complexity scales factorially with
the number of legs in Feynman-diagrammatic methods.
Much recent work has been devoted to the use of ``unitarity'', first
introduced for QCD loop calculations over 15 years
ago,\cite{Bern:2008ef} which, essentially, involves sewing together
tree-level amplitudes with specific kinematics in order to obtain the
coefficients of the loop integral.
Both Ma\^\i tre\cs\cite{MaitreProc} (for the Blackhat collaboration) and
Melnikov\cs\cite{MelnikovProc} (for the Rocket collaboration) reviewed
the amazing progress that has taken place in recent years (see also
ref.~\cs\cite{Bern:2008ef}), significant innovations including, among
many others, the use of complex momenta,\cite{Britto:2004nc} recursive
building up of the number of legs,\cite{Berger:2006ci} the
determination of the full analytic structure of loop integrands based
just on their numerical evaluation at a finite set of kinematic
points,\cite{Ossola:2006us} and extraction of results in $4+2\epsilon$
dimensions from computations in integer $D>4$
dimensions.\cite{Giele:2008ve}

\begin{figure}
  \centering
  \begin{minipage}[c]{0.3\linewidth}
    \includegraphics[width=\tw]{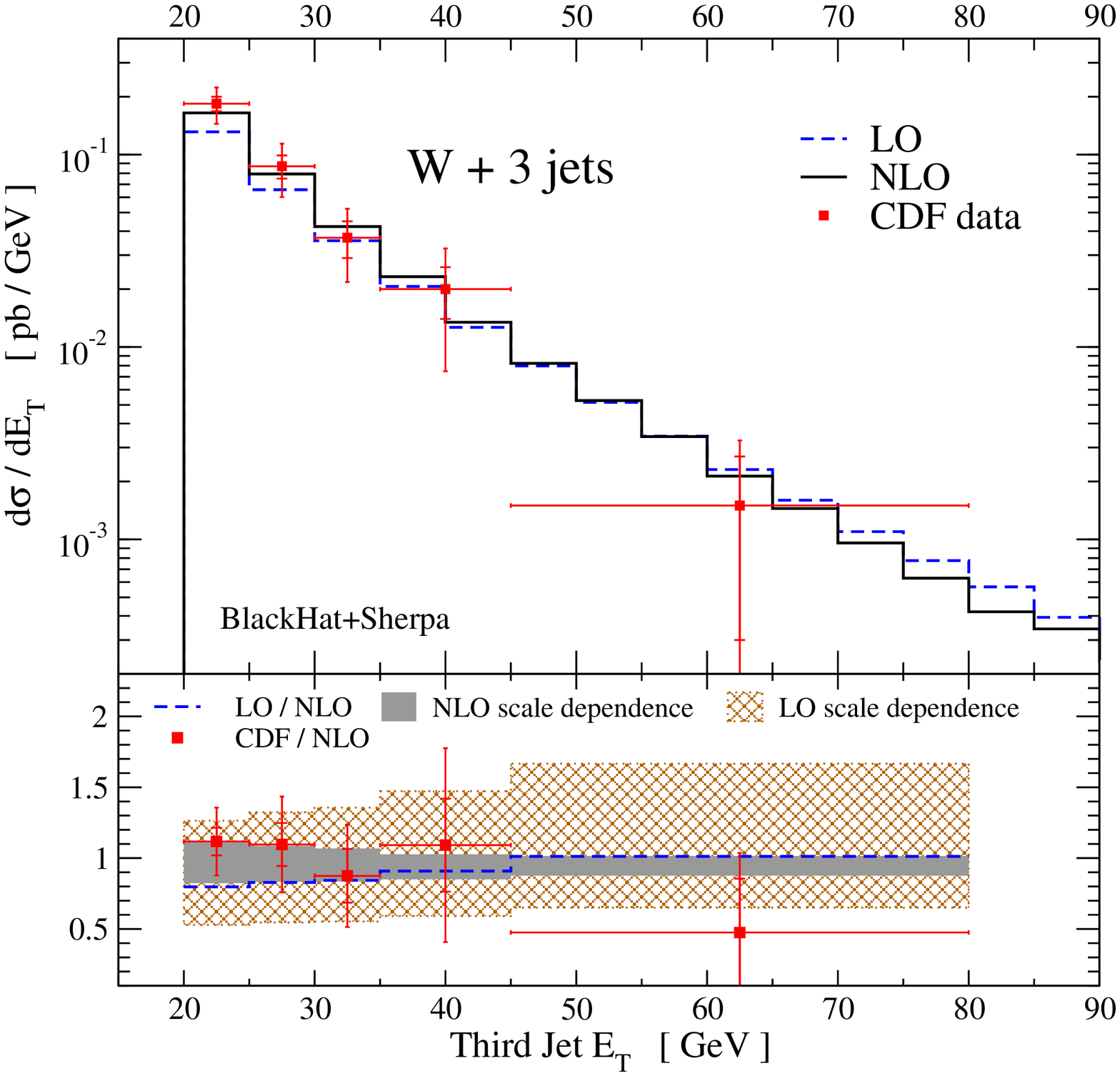}
  \end{minipage}
  \begin{minipage}[c]{0.3\linewidth}
    \includegraphics[width=\tw]{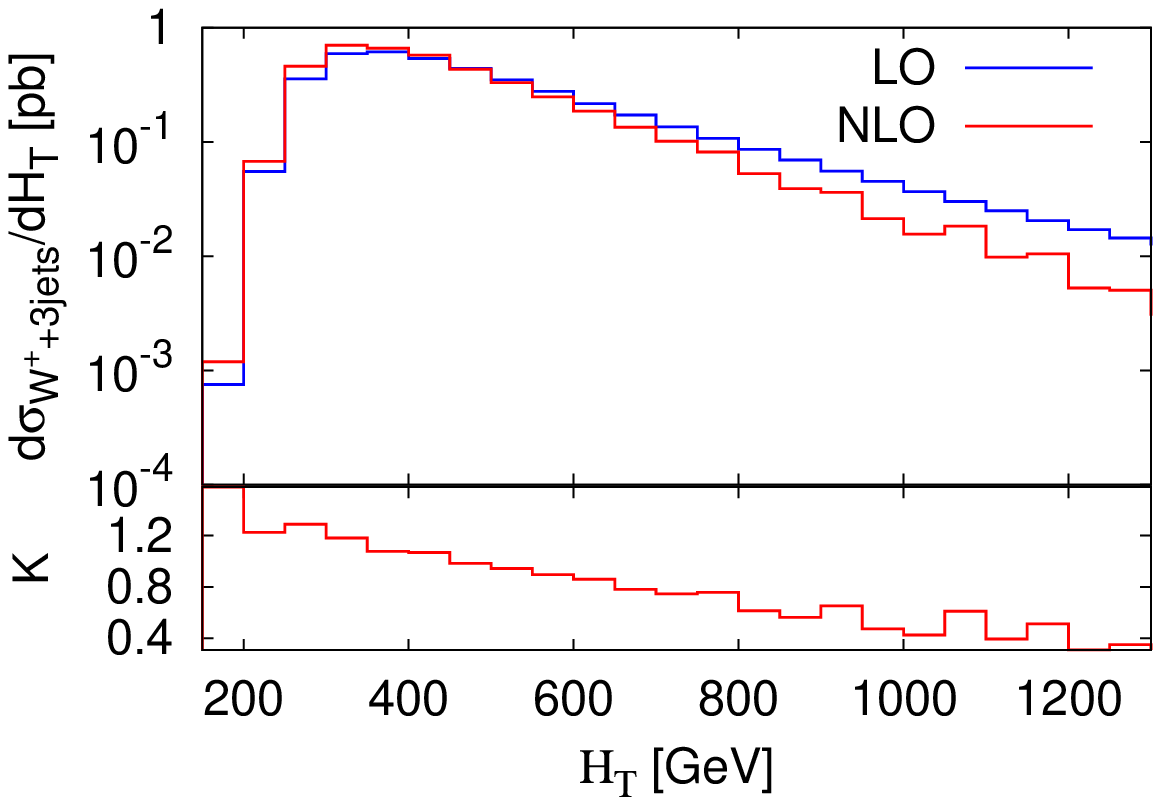}\vspace{-0.7em}\\
    \includegraphics[width=\tw]{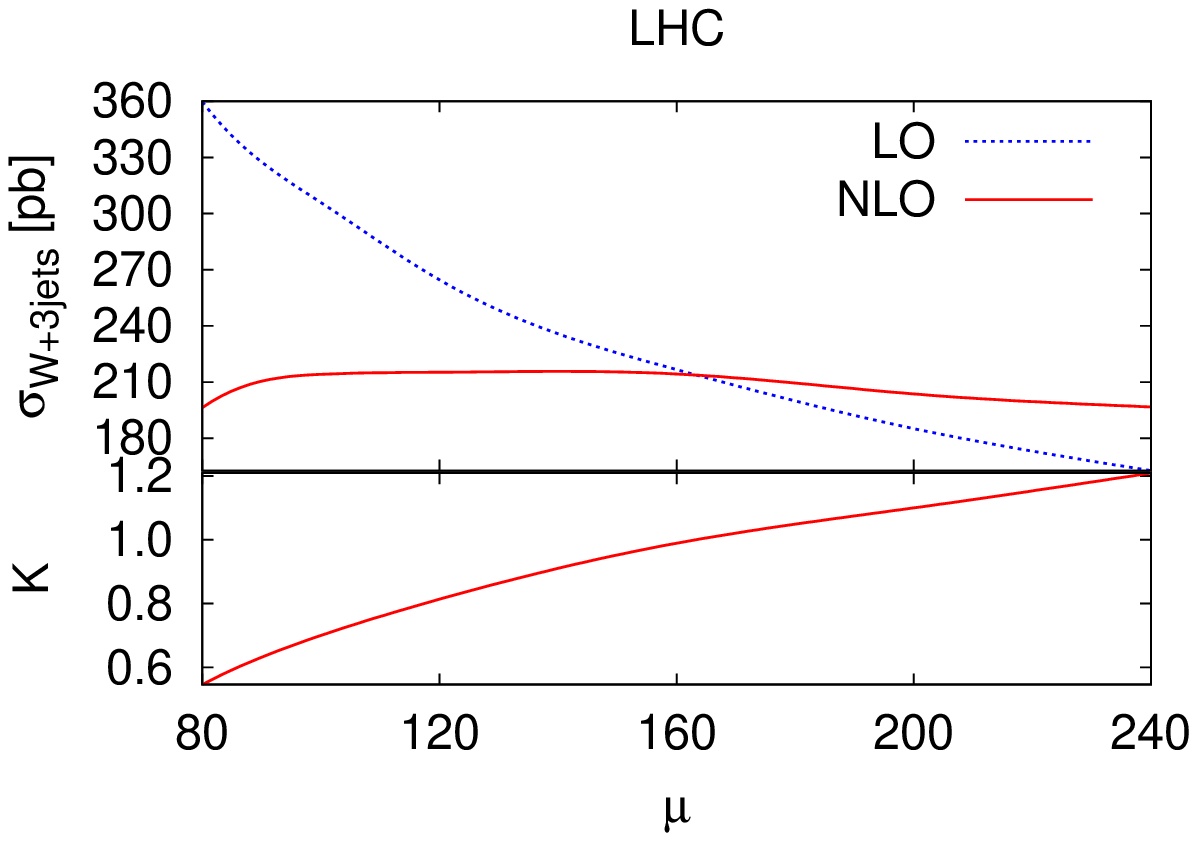}
  \end{minipage}
  \caption[]{Left: results\cs\cite{Berger:2009zg} from Blackhat for
    the $W+3$jet cross section as function of the $p_t$ of the softest
    of the 3 jets, compared to CDF data;\cite{Aaltonen:2007ip} right:
    predictions for the $H_T$ variable at the LHC from the Rocket
    program (top) and the scale dependence of the LHC cross section
    for $W+$3 jets with $p_t > 50\GeV$ (below).\cite{Ellis:2009zw}}
  \label{fig:W3jet-nlo}
\end{figure}

The power of these methods was conveyed through the list of 1-loop
amplitudes available in the ``Rocket'' program: all 1-loop $N$-gluon
scattering amplitudes,\cite{Giele:2008bc} $q\bar q$ + $N$-gluons,
$Wq\bar q + Ng$, $Wq\bar qq\bar q + Ng$, $t\bar t + Ng$ and $t\bar t
q\bar q+ Ng$.
In terms of phenomenological applications, it seems that $2\to4$ and
$2\to5$ processes are within realistic reach, at least in the
large-$N_c$ limit, and significant work is now being devoted to
the combination of the $2\to n$ 1-loop result with the $2\to n+1$
tree-level result (Blackhat uses Sherpa,\cite{Gleisberg:2008ta} Rocket
uses MCFM\cs\cite{Campbell:2002tg}).
Both groups showed first results for $pp\to W + 3$jets, one of the
major $2\to 4$ Les Houches processes (it's a major background to SUSY
searches). The results were in the large-$N_c$ limit (which should be
good to a few percent), and in the case of Rocket with just the
$Wq\bar qggg$ subprocess and without fermion loops (good to
$20-30\%$).
Some of them are reproduced in fig.~\ref{fig:W3jet-nlo}, including a
comparison to data\cs\cite{Aaltonen:2007ip} from the CDF
collaboration.\footnote{%
  A vexing issue here is that the data have been
  obtained with the JetClu jet algorithm, which is severely IR unsafe
  and causes even the LO perturbative prediction to be
  ill-defined.\cite{Salam:2008qq}
  To obtain finite NLO predictions, the Blackhat group instead used
  the SISCone jet algorithm.\cite{Salam:2007xv}
  It would probably be worth supplementing this with a calculation
  that uses an alternative such as anti-$k_t$,\cite{Cacciari:2008gp}
  insofar as JetClu is probably intermediate in its behaviour between
  anti-$k_t$ and SISCone, once one accounts for the all-orders
  perturbative and non-perturbative impact of the IR unsafety of
  JetClu.  }

These developments represent a major step forward and the start of a
new era in practical NLO calculations for the LHC and one can almost
certainly expect significant progress on the remaining technical
issues in the coming year or two.

\subsection{Not NLO}

Plain NLO calculations are not the only means available to us for
obtaining predictions at colliders and a number of varyingly related
methods were presented at this year's Moriond.

It can be useful to combine NLO predictions with a parton-shower Monte
Carlo simulation. White\cs\cite{WhiteProc} discussed this in the context
of MC@NLO\cs\cite{Frixione:2002ik,Frixione:2008yi} for $pp \to W t$
production. An issue that arises at NLO is the
appearance of the $pp\to W t\bar b$ process, which interferes with
non-resonant $t\bar t$ production. This is a non-trivial problem, and
to have a solution that allows $pp \to W t$ to be incorporated in
MC@NLO is a very useful development

Part of the interest of parton showers is that they resum logarithmically
enhanced terms to all orders. 
The best resummation precision is, however, to be obtained with
analytic calculations, which were discussed by
Ferrera\cs\cite{FerreraProc} for the $p_t$ distribution of a $Z/\gamma^*$
system. 
A context for this is that the $p_t$ distribution for the Higgs boson
(which is calculated in a similar way), is an important ingredient in
Higgs searches, and it is valuable to be able validate the
calculational framework for predicting this, which is very similar in
the Higgs and the Z cases.

Resummations may also be relevant in predicting the structure of
multi-jet events. Normally multi-jet predictions are based on
tree-level calculations, 
but it was pointed out by Andersen\cs\cite{Andersen} that in the case of
Higgs plus multijet production, it is technically difficult to obtain
exact predictions for multijet prediction. He thus discussed an
interesting approach\cs\cite{Andersen:2008gc} based on the
Fadin-Kuraev-Lipatov high-energy approximation,\cite{Kuraev:1976ge}
which compares well to exact tree-level calculations in the cases
where they are known. This is an interesting complement to normal
fixed order methods, in part also because it provides a natural way of
including virtual corrections. 
The relevance of the high-energy approximation was also emphasised by
Hautmann,\cite{Hautmann} because of the expected relevance at LHC of
configurations in which multiple emissions may have commensurate
transverse momenta (by default not included in parton shower Monte
Carlos).

Rather than trying to calculate all orders in some logarithmic
approximation, one can also try to obtain just one order further than
NLO, i.e. NNLO.
Work towards an efficient program for fully exclusive NNLO prediction
of $pp \to Z$ was presented by Ferrera.\cite{FerreraProc}
Theoretical developments were discussed by Heslop\cs\cite{HeslopProc} on the calculation
of two-loop diagrams (one of the ingredients of NNLO predictions) for a
theory related to QCD, ${\cal N}=4$ supersymmetric (SUSY) Yang-Mills
(YM) theory, specifically for maximal-helicity-violating (MHV)
amplitudes.
That large number of acronyms is indicative of how distant this is
from a general full QCD calculation. Yet the progress made is
impressive. In particular, Heslop discussed a conjecture that relates
gluon loop amplitudes to Wilson loops, and showed that if it holds,
then one can calculate all planar two-loop MHV n-gluon scattering
amplitudes in ${\cal N}=4$ SUSY-YM for any number of gluon legs
$n$.\cite{Anastasiou:2009kn} 
This, for two-loop diagrams, is analogous to the type of progress that
was being made 15 years ago for one-loop diagrams\cs\cite{Bern:1994zx}
and that recently has been playing a big role in NLO calculations, as
described in section~\ref{sec:NLO}.

In discussing perturbative predictions for high-energy colliders, it is
important to remember that non-perturbative effects can often be as
large as higher-orders of perturbation theory. 
This is especially true when it comes to the underlying event and
pileup at the LHC, and the simulation of these effects was discussed
by Pierog,\cite{PierogProc} in the context of the EPOS Monte Carlo program for
minimum-bias physics, including the question of how one can incorporate
constraints from cosmic-ray air showers in the modelling of
minimum-bias collisions.
%

\section{The Data--Theory interface}

Work at the interface between data and theory is crucial if we are to
make the best possible use of both. 
The topics that fell under this heading were rather varied.

\begin{figure}
  \centering
  \includegraphics[width=0.48\tw]{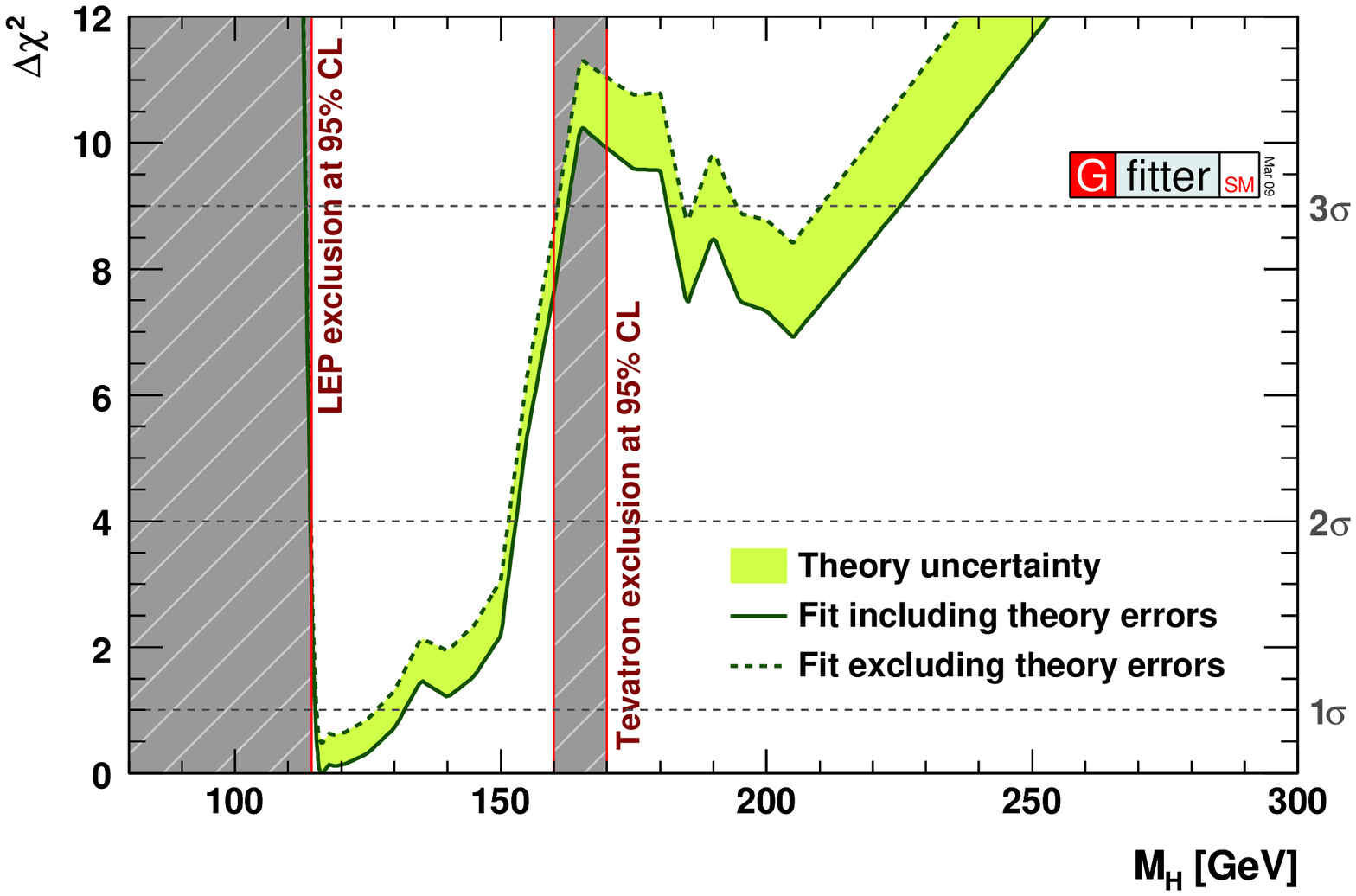}
  \includegraphics[width=0.48\tw]{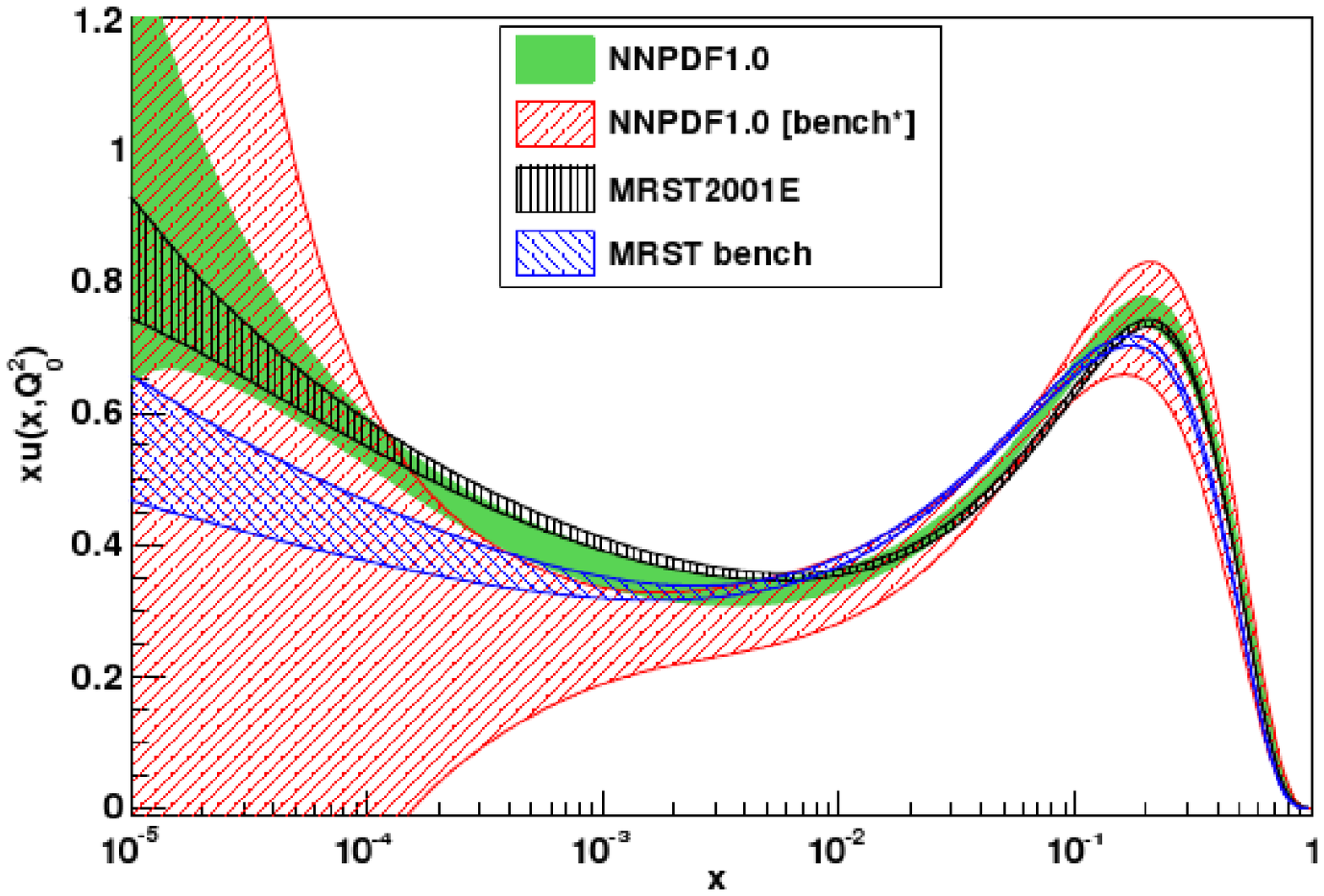}
  \caption[]{Left: the $\Delta \chi^2$ (with respect to the minimum)
    for the Gfitter\cs\cite{Flacher:2008zq} fit\cs\cite{StelzerProc} to the
    latest electroweak precision data, including direct
    searches. Right: fits for the up-quark distribution from the
    MRST/MSTW and NNPDF groups with full data, and ``benchmark''
    versions with a reduced dataset.\cite{Dittmar:2009ii}}
  \label{fig:EWfit}
\end{figure}

Stelzer\cs\cite{StelzerProc} discussed the Gfitter
project\cs\cite{Flacher:2008zq} for electroweak fits of the standard
model (and beyond). It can be seen as an alternative to a tool like
against Zfitter,\cite{Arbuzov:2005ma} and has also been validated
against it. 
Stelzer quoted a central value for the Higgs mass of
$83^{+30}_{-23}\GeV$, to be compared with that from the Tevatron's
electroweak fit of $m_H = 90^{+36}_{-27}\GeV$ (small details of the fit are
responsible for the difference in results).
Including the latest results for the direct Higgs searches gives $m_H
= 116^{+15.6}_{-1.3}\GeV$, with the $\chi^2$ as a function of $m_H$
shown in fig.~\ref{fig:EWfit} (left).

Still on the subject of using data to constrain theory,
Williams\cs\cite{WilliamsProc} discussed a program called
HiggsBounds,\cite{Bechtle:2008jh} which incorporates results of all
experimental Higgs-boson searches into a single package. 
The list of searches that are included in the program (too long to
reproduce here) makes for an impressive and valuable
collation of information.
It can be useful for testing new models (and there is a convenient web
interface), or even new standard-model cross sections, and Williams
illustrated how it had been used to show that a previous 95\%
exclusion limit on the Higgs boson from the Tevatron disappeared once
one used updated PDFs.

The question of PDFs is one that arises in many places, not
surprisingly given how crucial an input they are for Tevatron and LHC
studies.
A major issue in standard PDF fits is the determination of the
uncertainties. The two main groups, CTEQ and MSTW, both estimate them
using a $\delta \chi^2$ of order $50$. However reasonable the final
results, one can't but help feeling a little uncomfortable with this
choice.
A second issue is that standard fits use somewhat restricted
parametrisations, which may bias the final results. 
An approach that attempts to work around these issues was presented by
Del Debbio,\cite{DDProc} for the NNPDF
collaboration.\cite{Ball:2008by} 
One innovation is that they carry out individual fits to a large
number of Monte Carlo replica experiments so as to obtain an ensemble
of PDFs (i.e.\ a direct measure of uncertainties, without needing to
choose a $\delta \chi^2$ value).
Additionally, they use neural networks to provide bias-free
parametrisations of the PDFs. 
Fig.~\ref{fig:EWfit} (right) shows results of fits for the up-quark
distribution compared to MRST results. There are two fits each, one
using a full data set and the other a reduced ``benchmark'' data
set.\cite{Dittmar:2009ii}
Ideally, the original fit should be within the error band for the
benchmark fit, and the latter should have significantly larger errors
in the region lacking data. This is the case for NNPDF, but less so
for MRST, perhaps a consequence of the in-built parametrisation, which
provides a constrained extrapolation into the region with limited
data.
At the moment the NNPDF fit lacks heavy-quark effects and $p\bar p$
data, which limits its usefulness, however work is in progress to
resolve these issues.
Once this is done, it seems likely that the NNPDF approach will become
a serious competitor to the CTEQ and MSTW groups.

\begin{floatingfigure}[r]{0.4\textwidth}
  \centering
  \includegraphics[height=0.4\tw,angle=270]{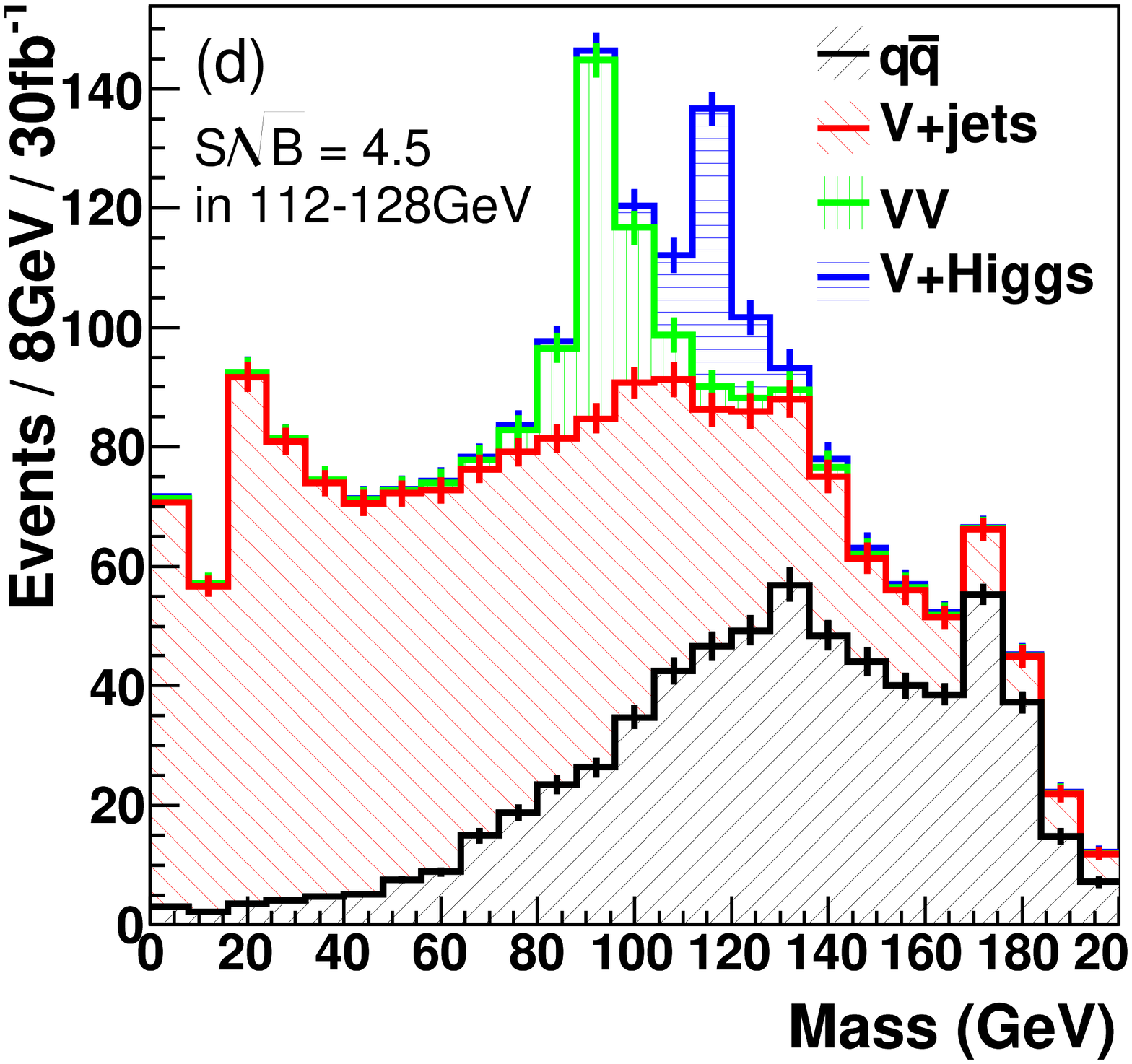}
  \caption[]{Distribution of the mass for tagged, high-$p_t$ $b\bar b$
    jets with appropriate substructure, in events also pass a leptonic
    and missing-$E_T$ (W/Z) cut, for a Higgs boson mass of
    $115\GeV$.\cite{Butterworth:2008iy}}
\label{fig:HiggsSearch}
\end{floatingfigure}
As well as using data to learn more about theory, one can also take
the reverse approach and ask how theoretical insight can be exploited
to better use data. 
This was illustrated in the talk by Rubin\cs\cite{Rubin:2009ft} about a
proposal for a new LHC search strategy for a light Higgs boson that
decays to $b\bar b$.\cite{Butterworth:2008iy} In an ATLAS
study\cs\cite{atlasphystdr} of the $pp \to HW$, $H\to b\bar b$ and $W\to
\l \nu$ channel, it was found that the signal to background ratio was
very low, as was the significance, with the signal only a tiny
perturbation close to a peak in the background distribution.
Rubin concentrated on the subset of $WH$ events where the $W$ and $H$
both have high transverse momenta. Though only a small fraction of WH
events are in this configuration, it turns out the fraction for the
background events is smaller still.
One challenge in then that the $H \to b\bar b$ decay is quite
collimated and the $b\bar b$ may end up in the same jet. However,
using a QCD-motivated dedicated subjet ID strategy, this issue can be
resolved and, based on Monte Carlo study, one expects that with
$30\,\mathrm{fb}^{-1}$ one could obtain a $4-5\sigma$ significance for
discovery of a $115\GeV$ Higgs boson at a $14\,$TeV LHC,
cf.~fig.~\ref{fig:HiggsSearch}.

The issue of very small signal to background ratios is one that is
common to many of the experimental analyses discussed at this year's
Moriond QCD. 
In nearly all the cases the analyses used a neural network (NN), or
some other multi-variate technique, to obtain a measure of how much a
given event is ``signal-like'' versus background-like, and then showed
the distribution for this measure as their main result.

In informal discussions during the conference, many people expressed
discomfort at this trend (myself included).
It is natural of course to seek to use the best tools at hand in order
to maximise one's chances of seeing a signal. 
However, ultimately, the aim is not merely to get the largest possible
value for some number such as $S/\sqrt{B}$, but, just as importantly,
to convince the reader/audience that one has actually seen (or
excluded) a signal.
From this point of view, neural networks may actually be a hindrance,
because they fail to communicate what it is about a certain set of
events that leads one to believe that they correspond to a signal.
One can perhaps mitigate this drawback, to some extent, by
showing the correlation between the neural network output and various
physical distributions for background and signal. 
However, a suggestion for a more general rule of thumb might be the
following: if a NN improves the signal significance by (say) 20\%
compared to a cut-based analysis, then one should also show the
latter, because it is likely to be just as convincing (if not more so).
If, instead, the NN improves the significance by a factor of two, then
this suggests that there is some underlying physical characteristic of
the signal that could be used also to improve a traditional analysis,
and one should figure that out.\footnote{Another way of saying this is
  that one ought not to excessively favour silicon-based neural
  networks over their carbon-based cousins.}

\section{Beyond the Standard Model}

Two kinds of ``New phenomena'' were discussed at this year's
Moriond. Those that relate to theories that we know well (QCD), but
that may have yet-to-be discovered exotic behaviour, for example in
heavy-ion collisions, as discussed below; and those that relate to
extensions of the standard model.

One of the issues with the most popular extension of the standard
model, supersymmetry (SUSY), is that of how this extra symmetry
between fermions and bosons gets broken. Various schemes exist in the
literature, such as gravity-mediated SUSY breaking, or
gauge-mediated SUSY breaking. Lalak\cs\cite{LalakProc} pointed out that,
contrary to standard assumptions, it is possible that real-world SUSY
breaking could be a mixture of these.

SUSY is far from being the only viable extension of the Standard
Model. Kanemura\cs\cite{KanemuraProc} discussed a specific
model\cs\cite{Aoki:2009vf} in which the dynamics of an extended Higgs
sector and TeV-scale right-handed neutrinos provide a framework for
neutrino oscillation, dark matter, and baryon asymmetry of the
Universe. In particular tiny, physical, neutrino masses are generated
at the three loop level, a singlet scalar field is a candidate of dark
matter, and a strong first-order phase transition is realised for
successful electroweak baryogenesis.

One of the most economical ideas for the explaining the electroweak
scale involves the idea that the Higgs is composite (a bit like pion),
with its mass generated by non-perturbative dynamics of a new QCD-like
theory, technicolour, but whose coupling grows strong near $1\TeV$
rather than near $1\GeV$.
Technicolour is often considered to be difficult to reconcile with
precision electroweak measurements, but as was discussed by
Brower,\cite{BrowerProc} this is based on calculations that assume
that technicolour is similar to QCD. If one instead supposes that
technicolour is only marginally similar to QCD (e.g.\ it has many more
active flavours, with a small $\beta$-function coefficient), then it
might be a rather different theory. Would this too then be excluded? The
only way of being sure would be through lattice calculations. In this
respect, the fact (cf.\ section~\ref{sec:lattice}) that we are
finally reaching an era of full control over the systematics of
lattice calculations means that we might also be to use them to
reliably address questions like this about technicolour.

\section{Heavy-Ion Collisions}

The key question in the study of heavy-ion collisions (HIC) is that of
whether we can understand the ``medium'' that is produced in such a
collision. 
Ways of addressing this question include direct modelling/calculation
of the medium, and the use of probes that traverse it to measure its
characteristics.

\begin{figure}
  \centering
  \includegraphics[width=0.48\tw]{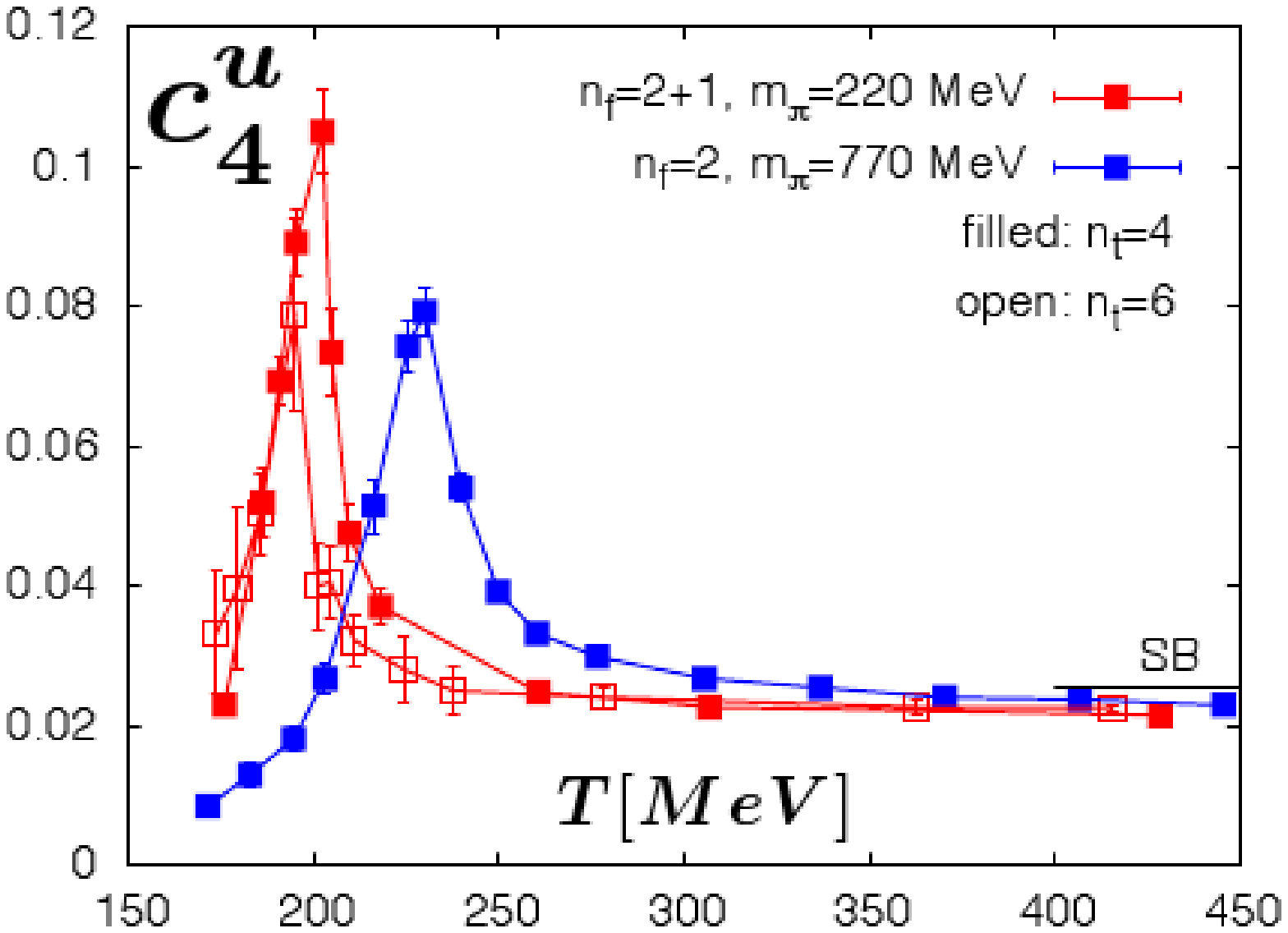}\hfill
  \includegraphics[width=0.48\tw]{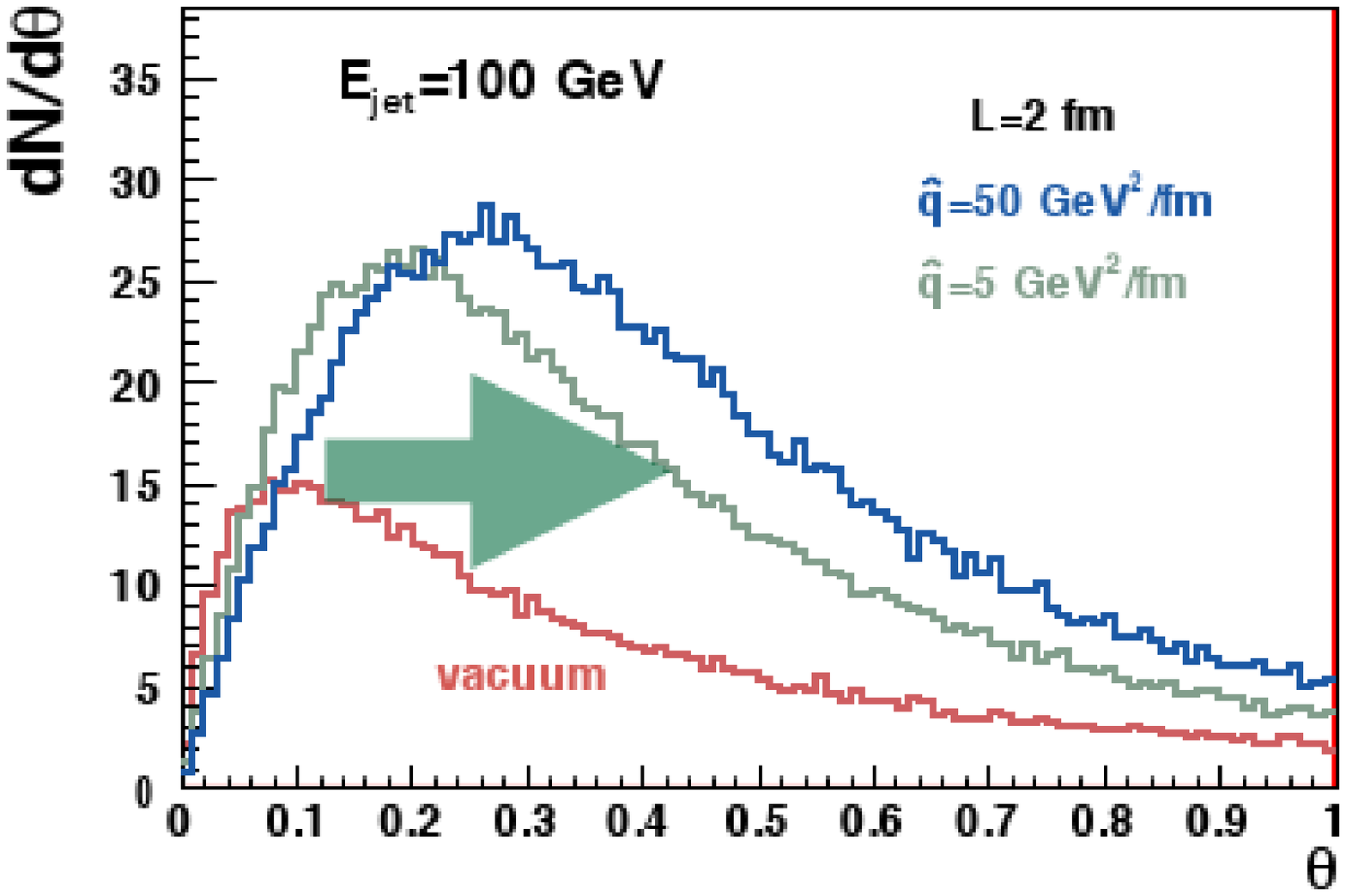}
  \caption[]{Left: the fourth-order coefficient of the Taylor
    expansion of the equation of state in lattice simulations of
    finite temperature QCD,\cite{SchmidtProc,Cheng:2008zh} with
    structure at $T\simeq200\MeV$ that is suggestive of a critical
    point. Right: the angular distribution of particles in that are
    produced from the showering of a 100\,GeV gluon in the vacuum, and
    in a medium with transport coefficient $\hat q = 5, 50
    \GeV^2/\mathrm{fm}$, as simulated with
    Q-Pythia.\cite{Armesto:2008qh}}
  \label{fig:critical-point}
\end{figure}

Direct calculation can be performed with lattice calculations at
finite temperature (albeit only for equilibrium, static media, i.e.\
an idealisation of what is to be found in true
HICs). Schmidt\cs\cite{SchmidtProc} described a lattice
calculation\cs\cite{Cheng:2008zh} of the equation of state for the
medium.
One interesting result was for the Taylor expansion of the pressure
as a function of the ratio of quark chemical potential $\mu_{uds}$ to
temperature $T$,
\begin{equation}
  \label{eq:eos}
  \frac{p}{T^4} = \frac{1}{VT^3} \ln Z(V,T,\mu_u, \mu_d,\mu_s)
  = \sum_{ijk} c_{i,j,k}^{u,d,s} 
  \left(\frac{\mu_u}{T}\right)^i
  \left(\frac{\mu_d}{T}\right)^j
  \left(\frac{\mu_s}{T}\right)^k
\end{equation}
The fourth-order Taylor coefficient, $c_4^u$ is shown as a function of
temperature in fig.~\ref{fig:critical-point}, and one sees a clear
peak near $200\MeV$. This, together with the behaviour of the other
expansion coefficients, hints at the existence of a critical point
at that temperature --- something that the experiments may
look for explicitly in their data.

Greiner\cs\cite{GreinerProc} discussed a microscopic approach to the
quark-gluon plasma, the ``Boltzmann Approach of MultiParton
Scatterings'' (BAMPS), a transport algorithm that solves the Boltzmann
equations for on-shell partons with perturbative QCD interactions,
essentially including a $2\to2$ scattering term and a $2\to 3$ term,
the latter being important for thermalisation.
This gives a very good description of the elliptic flow,
$v_2$ (a non-trivial achievement),\cite{Xu:2008av}  but significantly
oversuppresses high-$p_t$ particle production, giving
$R_{AA}\simeq0.05$ as opposed to the experimental value of
$R_{AA}\simeq0.2$.
So, while a microscopic description can provide much insight, it seems
that it remains a challenge to describe the entire body of data.

A complementary approach to the calculation of medium properties is to
investigate the influence of the medium on probes that traverse it, so
as to obtain a measure of its properties (albeit a somewhat indirect
one). 

One classic probe is the $J/\psi$, on the grounds that in a hot medium
it would ``melt'' and so its production would be
suppressed. Ferreiro\cs\cite{FerreiroProc} pointed out that to make sense
of AA data, it is important\cs\cite{Bravina:2009gz} to understand not
only the high-temperature effects, but also phenomena such as nuclear
shadowing that occur even in cold nuclear matter.
Another ``early-time'' probe was discussed by
Kerbikov,\cite{KerbikovProc} specifically $\pi\Xi$ correlations,
of interest because models predict an early decoupling of multi-strange hadrons
like the $\Xi$.

A probe that has seen very extensive study in recent years is
a hard parton that traverses the medium. The main indicator that has
been discussed so far is the amount of energy lost during this
traversal, which has been modelled in terms of medium-enhanced
radiative energy loss, as well as collisional energy loss.

Zakharov\cs\cite{ZakharovProc} discussed an additional source of energy
loss. He argued that certain models of the quark-gluon plasma, such as
``anomalous viscosity''\cs\cite{Asakawa:2006tc} imply the presence of
chromo-magnetic fields that are sufficient to induce substantial
synchrotron radiation from a gluon that goes through
them.\cite{Zakharov:2008uk} He estimated that synchrotron energy loss
should be of similar magnitude to collisional energy loss, and each of
them about $25\%$ of radiative energy loss.

The fact that many mechanisms may contribute to parton energy loss
motivates more exclusive studies, which look not just at leading
particle spectra, but the properties of particle and energy flow in
the vicinity of a leading particle.
To help interpret such studies, it is essential to have
more exclusive modelling tools, such as ``medium-aware'' Monte Carlo
generators. Salgado\cs\cite{SalgadoProc} discussed a modification of
Pythia,\cite{Sjostrand:2003wg} Q-Pythia,\cite{Armesto:2008qh} that
incorporates an additional medium-induced gluon emission term in the
parton shower.
Fig.~\ref{fig:critical-point} (right) shows the angular distribution
of particles emitted from a $100\GeV$ gluon, both in the vacuum and in
a medium, and illustrates the significant differences that are to be
seen, both in multiplicity and in typical angle.
This kind of tool promises to be very useful, both in testing the
underlying modelling, and in designing experimental analyses to
further probe the mechanism of jet quenching.

The final talk of the conference, by Warringa,\cite{WarringaProc}
discussed the effects of topological charge change in heavy-ion
collisions.\cite{Kharzeev:2007jp} He argued for the following chain of
events: 1) there are topological charge fluctuations in the hot medium
(like instantons); 2) that topological charge fluctuations induce
fluctuations in chirality, e.g.\ more right-handed quarks and
antiquarks (i.e.\ with the spin aligned along the direction of motion)
than left-handed ones; 3) that if there is a (QED) magnetic field in
the medium, this will orient the spins of the $u_L, \bar d_L, u_R,
\bar d_R$ quarks parallel (and the others anti-parallel) to the
direction of the magnetic field; 4) that together with a fluctuation
in chirality, (say more R), this will lead to $u_R \bar d_R$ (positive
charge) moving in the direction of the field, and $\bar u_R d_R$
(negative charge) moving in the opposite direction; 5) that in a
non-central AA collision there is a (QED) magnetic field,
perpendicular to the reaction plane, generated as the two charged
nuclei go past each other, and therefore that this orients net charge
flow along a direction perpendicular to the reaction plane; 6) that
this can be seen in AA collisions, by plotting a variable $\langle
\cos(\phi_i^{\pm} + \phi_j^{\pm,\mp} - 2\Psi_{RP})\rangle$ (where
$\Psi_{RP}$ is the angle of the reaction plane) as a function of
centrality. 
Remarkably, a preliminary STAR measurement\cs\cite{Caines:2009yu} shows
that this quantity becomes significantly different from zero (in the
direction expected) for the least central collisions, precisely those
that are expected to have the strongest magnetic field.
Given how long the theoretical community has been discussing
topological charge in SU(N) theories, this is a very interesting development.
One can only look forward to further cross-checks, and one interesting
one would for example be the experimental verification of appropriate
scaling with nuclear charge $Z$ for constant $A$.


\section{Acknowledgements}

This summary would not have been possible without the many patient
explanations that numerous Moriond speakers were kind enough to
provide me with, both about their own talks and about the more general
questions of their sub-fields.
That said, the inevitable errors, misunderstandings or
misappreciations that have made their way into these proceedings are
entirely mine.

I am also grateful to Luigi Del Debbio and Carlos Salgado for helpful
comments on the manuscript, as well as to Al Mueller for a useful
remark.

Finally, it would not possible to close without also thanking the Moriond
organisers --- for the invitation to give this summary (and financial
support), for their excellent organisation of the conference, and for
the ``Spirit of Moriond'' that has made this series of meetings such a
success over the several decades of its existence.

%

\end{document}